# Rich diversity of crystallographic phase formation in 2D $Re_x:Mo_{1-x}S_2$ ($x < 0.5$) alloy

H. Sharona,[1,2] B. Vishal,[1,2] U. Bhat,[1,2] A. Paul,[1,2] A. Mukherjee,[1,2] S. C. Sarma,[3,4] S. C. Peter,[3,4] and R. Datta[1,2] *

[1]*International Centre for Materials Science*, [2]*Chemistry and Physics of Materials Unit, Jawaharlal Nehru Centre for Advanced Scientific Research, Bangalore 560064, India.*

[3]*New Chemistry Unit*, [4]*School of Advanced Materials, Jawaharlal Nehru Centre for Advanced Scientific Research, Bangalore 560064, India.*

**Abstract**

We report on the observation of rich variety of crystallographic phase formation in $Re_xMo_{1-x}S_2$ alloy for $x < 0.5$. For $x < 0.23$, no low dimensional super-structural modulation is observed and inter-cation hybridization remains discrete forming dimers to tetramers with increasing Re concentration. For $x > 0.23$, super-strutural modulaton is observed. Depending on the Re concentrations ($x = 0.23, 0.32, 0.38$ and $0.45$) and its distributions, various types of cation hybridization results in rich variety of low dimensional super-structural modulation as directly revealed by high resolution transmission electron microscopy. These layered alloy system may be useful for various energy and novel device applications.

Corresponding author e-mail: *ranjan@jncasr.ac.in*

## I. Introduction

Atomically thin transition metal dichalcogenides (TMDs) e.g., MoS$_2$, WS$_2$, ReS$_2$ etc. have attracted immense research attention in recent times due to their unique electronic properties in monolayer, few layers, and heterostructure forms.[1-12] These layered TMDs exhibit various structural polytypes with distinct properties which are useful in many novel devices and energy applications.[11,13-17] For example, the most stable form of monolayer MoS$_2$ is 1H (trigonal prismatic, space group $P\bar{6}3/mmc$) which is a semiconductor with fundamental direct band gap of 1.88 eV [$K \rightarrow K$]. In bulk form, MoS$_2$ is an indirect band gap semiconductor with a band gap of 1.42 eV [$\Gamma \rightarrow \Lambda$].[18] Metastable 1T form of MoS$_2$ is metallic in nature which is useful for charge extraction and the modulated 1T$_d$ semiconducting form is found to be suitable for hydrogen evolution reaction (HER) and supercapacitor application.[19,20] Theoretical calculations of various other possible phases such as 1T′ ($\sqrt{3}a \times a$) is predicted to exhibit non-trivial topological states[21] and the trimeric ($\sqrt{3}a \times \sqrt{3}a$) phase is predicted to be a thinnest ferroelectric material.[22] Since, each phase has their own unique properties, various efforts have been made to stabilise and engineer such phases.[23-26] In this context it is already reported that Li intercalation stabilizes $2a \times 2a$ (1T$_d$) polytype whereas, K intercalation stabilizes $\sqrt{3}a \times \sqrt{3}a$ ($x \approx 0.3$) and $2a \times 2a$ (1T$_d$) ($x \leq 0.3$) polytypes depending on the K concentration.[27,28]

On the other hand, ReS$_2$ belongs to Gr.-VII TMDs (space Gr. $P\bar{1}$) which has a distorted octahedral structure and often termed as 1T$_d$.[29] In 1T$_d$ phase, four Re atoms hybridizes to form super-structure. The origin of this type of periodic modulation lies at low dimensional Peierls distortions which is favoured if the strain energy associated with the modulation is compensated by opening up a band gap at the Fermi level.[30] ReS$_2$ retains its fundamental direct band gap in more than monolayer form unlike $MoS_2$ and this is attributed to its weak interlayer electronic and vibrational coupling.[31] The band gap of $ReS_2$ is 1.52 and 1.42 eV in monolayer and bulk form, respectively which is advantageous over $MoS_2$ in fabricating various thin electronic devices.[19]

As already mentioned that the stabilization of various structural polytypes of layered TMDs have been reported to exploit their unique properties. Similarly, alloys between different TMDs have been explored to tailor the structure and properties.[32-37] In this context, experimental study in $Mo_{1-x}Re_xS_2$ alloy system revealed two different polytypes at $x = 0.5$ and a trimeric polytype at $x = 0.25$.[32] The density functional theory (DFT) based calculation indicated that the $Mo_{0.5}Re_{0.5}S_2$ alloy may be a suitable candidate for nanoscale switching application involving transformation between semiconducting to conducting state with a minute perturbation of ~ 90 meV and shown to be an efficient catalyst for HER.[32-38] The effect of Re substitution in $MoS_2$ lattice on the structure and property has been reported by few groups.[24,38-43] Re prefer to substitute Mo lattice sites[43] and act as a $n$-type dopant by creating a shallow and deep defect level in bulk and monolayer form, respectively.[44,45] HRTEM and HAADF imaging revealed various Re positions in $MoS_2$ lattice: isolated dopant, dimer, trimer, tetramer and pathways to phase transition from 2H to $1T_d$ structure of the alloy depending on Re concentrations.[32] A structural cross over from 1H to 1T polytype at $x = 50\%$ has been predicted by theoretical calculation.[32] The detailed dynamics involved in such a phase transition can be found in a report where it was shown that the Re atom act as a nucleation site.[43] Subsequent report shows the composition dependent structure of $Mo_{1-x}Re_xS_2$ ($0 < x < 0.2$) alloy by X-ray photoelectron spectroscopy where Re-Re undergoes dimerization at $x = 0.05$ and tetramerization with further increase in Re concentration.[39]

In the present report, we describe the rich variety of crystallographic phase formation for Re concentration less than 50% as observed directly under HRTEM. No periodic super-structural modulation is observed for $x < 0.23$, but discrete dimers and tetramers are observed to be formed depending on Re concentration. For $x$ = 0.23, 0.32, 0.38, and 0.45, different types of super-structural modulation are observed and the asymmetry in modulation vector increases with increasing Re concentration. These are with supercells $2a \times 2a$ for $x$ = 0.23, 0.32 and $2a \times a$, $\sqrt{3}a \times a$ for $x$ = 0.38 and 0.45, respectively. Density functional theory (DFT) based calculations suggests the possibility of various types of cation-cation hybridization in this system which occurs not only between Re atoms but also between Re-Mo atoms both in the 1H and $1T_d$ configuration. A fine sampling of composition in theoretical calculation reveals the structural cross over between 1H and $1T_d$ to be at ~37.5%. The interaction between Re-Mo atoms other than expected interaction between Re atoms is a unique finding of the present investigation. The observed diversity of structural phases may provide an opportunity to exploit this alloy system in various novel device and energy application.

### II.A. Material synthesis and experimental methods

In short, the synthesis process involved first mixing the Mo (powder, 99.9%, Sigma Aldrich), Re (1.00 mm diameter wire, 99.97%, Alfa Aesar) and S (powder-325 mesh, 99.5%, Alfa Aeser) in ratio of 3:1:8 and then placing it in a sealed in a quartz tube ($10^{-5}$ mbar). The sealed tube is then heated to 200 °C with a heating rate of 10 °C/h followed by annealing at the same temperature for two hours. After this the temperature is increased to 900 °C with a heating rate of 20 °C/h followed by annealing at the same temperature for 120 hours. The quartz tube containing powders is then cooled to room temperature under air and then sample is collected for various study.[32]

All the HRTEM images are recorded in an aberration corrected FEI TITAN 80-300 keV transmission electron microscope. To exfoliate monolayer of specimens, powder sample is first probe sonicated for 45 minutes with a probe of tip size of 3 mm and then dropped on a holey

Carbon grid. Composition analysis is carried out by EDS to identify areas with different Re concentration for high resolution imaging.

### II.B. Calculation methods

Electronic structure calculations are performed using all electron density functional theory based WEIN2k code.[46] To solve Kohn-Sham equation this code utilises a full potential linearized augmented plane wave (FLPAW) + local orbital as basis set. The calculations are performed for two different polytypes of monolayer alloy $Mo_{1-x}Re_xS_2$ with different $x = 0.625$, 0.125, 0.1875, 0.25, 0.3125, 0.375, 0.4375 and 0.50. These polytypes are 1H and various $1T_d$ configurations. The most stable configurations in case of pristine $MoS_2$ and $ReS_2$ are 1H and $1T_d$, respectively. The alloy of 1H configuration is created by using a supercell of 4×4×1 from pristine $MoS_2$ with different number of Re substitution at Mo site to obtain the desire concentrations. On the other hand, for $1T_d$ structure with a $2a \times 2a$ superstructure modulation a 2×2×1 supercell is considered. To optimize lattice parameter, generalised gradient approximation (GGA) with Perdew-Burke-Ernzerhof (PBE) exchange correlation functional are used. The Self consistence field (SCF) cycles are then followed with energy, charge and force convergence criteria below 0.0001 Ry, 0.0001e and 1mRy/au respectively. The optimized in-plane lattice parameters of 1H $MoS_2$ is $a = b = 3.1917$ Å and for $1T_d$ $ReS_2$ is $a = 6.4308$ Å , $b = 6.4912$ Å and $\gamma = 119.03°$, all alloy calculation was started with these lattice parameters for respective structural configurations. The muffin tin radius for atoms are chosen in such way to avoid overlap between each other and -6 Ry energy separation of core and valence states is chosen. The $R_{mt}K_{max}$ parameter is set to 7 where $R_{mt}$ is the muffin tin radius of the smallest atom (Sulphur). The 1000 k points are used in first Brillouin zone during SCF cycle. To study the stability of different phases cohesive energy values are extracted for various alloy phases.

### III. Results and Discussion

The rich variety of structures as recorded in HRTEM images is described first in succession with increasing Re concentration. The observed structural variations can be classified into two broad categories based on the cation-cation hybridization: (1) short range interaction and (2) periodic super-structural modulation. Short range interaction depicts local inter-cation interactions involving two or few atoms with shorter interatomic bond length compared to the surroundings. This is observed for alloys with $x < 23\%$ where the global structure remains as 1H-polytype. The periodic super-structural modulation involving inter cation hybridization is observed for $x > 23\%$. The periodic super-structural modulation changes systematically in terms of asymmetry in modulation vectors along different primary lattice vector directions with increasing Re concentrations in the lattice. The HRTEM images, FFT pattern, and theoretically optimized schematic structures along <0001> Z.A for both MoS$_2$ and ReS$_2$ are given in Fig. 1. The 1H structure of MoS$_2$ has hexagonal geometry where Mo-Mo and S-S separation on the projected (0001) plane is 3.2 Å (Fig. 1(a)). In case of ReS$_2$, four Re Atoms hybridized to form Re$_4$ cluster leading to a quasi 2D chain-like $(2a \times 2a)$ superstructure (Fig. 1(b)). The superlattice spots corresponding to the superstructure of ReS$_2$ is shown in Fig. 1(d) and one such spot is marked with the red circle. Various Re-Re bond-lengths are outlined. The observed superstructures of the alloy system can be described in terms of deviation from the ideal MoS$_2$ and ReS$_2$ structures. Now to define the super-structures for the alloy system a unit cell consisting of four rhombuses is sketched with cations at its corners (Fig. 1(e)). This can be used to define various super-structural configurations in terms of inter atomic lattice vectors and angle subtended by them. The cation-cation hybridization can occur along three directions i.e. $a_1$, $a_2$ and $a_3$ with different magnitudes of modulation vectors. The observed modulation for various alloy systems is significantly different compared to ReS$_2$. To distinguish such modulation from ReS$_2$, we define a parameter $\Delta a_n = a'_n - a_n$ where $n = 1, 2, 3$ and the magnitude of $\Delta a_n$ indicates strength of a modulation. In case of ReS$_2$, $\Delta a_1 = 0.6$ Å , $\Delta a_2 =$ 1 Å, and $\Delta a_3 = 0.8$ Å, showing stronger modulation is along the diagonal followed by the two sides of the rhombus. In case of MoS$_2$, all the $\Delta a$ values equal to zero indicating absence of any superstructure.

### III.A. Mo$_{1-x}$Re$_x$S$_2$ alloy with $x < 20\%$

As already mentioned, that the monolayer of MoS$_2$ has lattice parameters of $a = b = 3.2$ Å and $\alpha = 60°$ exhibiting hexagonal symmetry. Though there are few reports on the nature of Re substitution in MoS$_2$ lattice either at lower concentration ($< 1\%$),[44,47] and intermediate concentration (25, 50%),[32,39] however for various other intermediate concentrations detailed aspects on the local structure and possible electronic properties are not available. In the present investigation, for $x = 1 - 17\%$, HRTEM images reveals local hybridization between two cations forming a dimer and an increase in population of such local dimers with increasing Re concentration. The Figure 2 (a) & (e) are the HRTEM image of Mo$_{0.99}$Re$_{0.01}$S$_2$ alloy and example line scan showing various cation-cation distance, respectively. Most of the inter-cation distances are ~3.2 Å, which is typical of Mo-Mo interatomic distance with local random values of 2.79 Å as indicated in the Fig. 2(a). This local short inter-atomic distance is due to dimer formation involving pair of cations. The density of such random dimers increases with Re concentration of 7, 14, and 17 % (Fig. 2(b)-(d)). Local clustering involving more than two cations can be observed for $x = 17\%$ and is indicated in the image (Fig. 2(d)). These are local tetramer type modulation having average lattice parameters: $a_1 = 2.6/2.8$ Å, $a_1' = 3.1$, $a_2 = 2.8/2.9$ Å, $a_2' = 2.9/3.1$ Å, $\alpha_1 = 58°$ and $\alpha_2 = 55°$. The cluster formation in this case can be between Re and Mo atoms other than expected interaction between Re atoms as confirmed from the intensity of line scan and later by theoretical calculation.

### III.B. Mo$_{1-x}$Re$_x$S alloy with $x > 20\%$

Formation of various super-structures due to cation-cation hybridization is observed to form for $x > 20\%$. However, the magnitude of super-structure modulation vectors systematically becomes asymmetric with increasing Re concentration from 23 to 45%. The lattice vectors corresponding to various superstructures are summarized in Table I. The Fig 3 (a) & (b) are HRTEM image and corresponding FFT pattern for $x = 23\%$ alloy, respectively. From FFT pattern superlattice vectors corresponding to $1/2 < 11\bar{2}0 >$ is marked with a red circle. There is a large dispersion in lattice parameters and the average values are $a_1 = 2.88$ (±0.18) Å, $a_3 = 3.17$ (±0.16) Å, and $a_2 = 3.01$ (±0.12) Å as evaluated from the HRTEM images. The angle between the lattice vectors deviates from the ideal 60° angle corresponding to hexagonal structure (Table II). As the inter tetramer cluster distances are almost close to each other therefore from HRTEM image no clear asymmetric separation of clusters in terms of continuous dark channel contrast is observed which otherwise gets prominent with increasing Re concentration. The observed large variation in lattice parameters is marked with the colour lines representing different magnitudes (Fig. 3 (e)). The average superstructure in this case can be approximated by a $2a \times 2a$ unit cell (inset Fig. 3 (d)).

The Fig. 3(f)-(j) are the structural details for alloy with $x = 32\%$. Clear super structure modulation can be observed from the HRTEM image (Fig. 3(f)). Four cations together forming tetramer cluster due to hybridization. Dark channels are visible at inter cation clusters regions. The dark channels are more prominent and continuous along X direction compared to Y direction as indicated in the figure. Along Y direction, at some local places the width of the dark channel is varying suggesting local change in inter-cluster separation. This can occur if there are different types of cation participating into hybridization. In the present case between both Re-Mo and Re-Re and it is supported by theoretical calculation. Corresponding FFT image is given in Fig. 3(g). The asymmetry in inter-cluster separation is also captured in terms of brightness of superlattice spots (marked with blue and red colour circles) which is different along different directions. There are two pairs of bright spots marked by blue circles and one pair of weak spots marked by red circle in FFT (Fig. 3(g)). The superlattice spots in FFT pattern is stronger compared to the 23% case. The average super structure modulation is 2a×2a (Fig. 3(h)). The average structural parameters are given in Fig. 3(i). In this case the variation in lattice vectors is observed as well and is indicated with the colour lines in Fig. 3(j). The details of the structure for $x = 38\%$ is given in Fig. 3(k)-(o). In this case, inter-cluster separation or magnitude of modulation along one direction is more compared to other directions. The asymmetry or anisotropy is clearly visible in the FFT pattern (Fig. 3(l)) and is marked with the blue circle along $\bar{1}100$ for brighter spots and red circles for less bright spot along $01\bar{1}0$ and $\bar{1}010$ directions. The average super-structure is 2a×a and the structural parameters are given in Fig. 3(n). Local variation of inter cluster distance can be measured from the HRTEM image and is shown in Fig. 3(o). The asymmetry in direction of structural modulation further increases for alloy $x = 45\%$ and the structure appears as one-dimensional chain like modulation or zig-zag type [Fig. 3(p)-(t)]. The average supercell is of $\sqrt{3}a \times a$. The variation in structural parameters is reduced in this case and is shown in Fig. 3(t).

As already pointed out that the origin in rich variation of super-structural modulation lies at the inter-cationic interactions which is found to be diverse in nature. This suggests the possibility of interaction interactions between Mo-Mo, Mo-Re besides expected Re-Re interactions. To reinforce this anticipation, we have carried out DFT calculation for example alloy composition $x = 12.5$ and 25 % in 1H and 1T$_d$ structural configurations, respectively. The schematics for Re$_x$Mo$_{1-x}$S$_2$ alloy for $x = 12.5\%$ in 2H configuration with two Re atoms with two different configurations are given in Fig. 4. In the first configuration, two Re atoms are separated by Mo atom and in the second configuration two Re atoms are next to each other which is marginally stable by 19.5 meV compared to the first configuration. Both the configurations are energetically more favourable compared to random distributions of Re.

Presence of interaction can be observed between one Re with two Mo atoms and two Re and one Mo atom for the configurations 1 & 2, respectively. This supports the formation of dimer in 1H configuration of the alloy lattice. This result is similar to the Svetlana et al but in that case no inter cation interaction types is considered.[39] In general, Re atoms in lattice tend to hybridises with nearest Mo atoms to form dimers with bond length of 2.9 Å locally. The interaction distance in case of Re-Re dimer is 2.8 Å. The Re-Mo dimer formation occurs due to electron transfer from 5d orbital of Re to nearest Mo 4d orbitals Fig. 4(c) and Re-Re dimer are formed by charge sharing of 5d orbitals of Re forming a covalent bonding Fig. 4(d). The valence charge density plot for the both the cases are shown in Fig 4. (c) & (d) where charge sharing between interacting cations is clearly visible. In a paper published by Brandao et al[47] where it was shown by electron paramagnetic resonance technique that in $MoS_2$, intrinsic Re impurities found to introduces local site symmetry reduction and anisotropic electronic delocalisation favouring inter cation interaction.

The Fig. 5(a)-(e) are the schematic of alloy structures considered for the theoretical calculation in $1T_d$ configuration for $x = 25\%$ with different distribution of Re and Mo atoms in the lattice. The schematics are presented in the order with decreasing stability and corresponding energy and lattice parameters can be found in Table III. The most stable configuration in the list is structure (*a*) and the energy difference between structure (*a*) and (*b*) to (*e*) are -2.85, -14.97, -15.58, -189.01 meV, respectively. In the most stable configuration of (*a*), there are two similar zig-zag chains of cations consisting of alternate Re and Mo atoms (Fig. 5(a)). The cations are interacting between them along the bonding lines as shown in the bottom row. In structure (b), Re is along the direction as marked by black arrow and there is a modulation in inter Re distance though there is no interaction between them. Distributing Re preferentially from two chains to single chain increases the energy of systems by 2.85 meV which is marginal (Fig. 5(b)). In Fig. 5(c) there is a broken zig-zag chain structure consisting of Mo tringles and alternate interaction between three Mo and two Re atoms. There is a slight energy cost for structure (c) by ~12 meV compared to structure (b). Four atoms cation clusters forming periodic structural modulation of Mo and Re atoms is shown in Fig. 5(d). The structure (e) where all the Re atoms are close to each other is energetically least favourable i.e. 189.01 meV compared to structure (a). From the above calculations it is clear that the interaction between different types of cations are possible and even preferable over expected interactions between Re atoms only.

In this context Svetlana et al showed the formation of Re-dimer in the MoS$_2$ lattice for $x = 5\%$ by X-ray photo absorption spectroscopy (XPS).[39] An increase in Re concentration, rhombus like cluster corresponding to Re tetramer was observed to form. The HRTEM image shows local tetramer formation for $x = 15\%$ and the DFT calculation for 10% alloy affirms higher stability of tetramer cluster compared to dimer and random Re distribution in 1H lattice of MoS$_2$. However, in this case there is no diverse types of inter-cation interaction considered for the calculation. In another report, an enhanced HER activity was reported for MoS$_2$ alloyed with Re at 50% in 1T$_d$ configuration.[38] By HRTEM random distribution of Re at 5% and tetramer cluster at 55% in the MoS$_2$ lattice in DT ($2a \times 2a$) configuration was shown. First principle calculation shows the stability of DT phase is more than 2H for $x = 50\%$ particularly at wider channel regions of the DT structure. However, for $x$ higher than 50%, HER activity was reported to decrease due to reduced density of states at the Fermi level.

The pristine monolayer 1H MoS$_2$ has direct band gap of 1.7 eV at $K$ point[18] with valence band maxima (VBM) composed of Mo $4d_{x^2-y^2, xy}$ and S $3p_{x,y}$ orbitals and conduction band minima (CBM) consists of Mo $4d_{z^2}$ and S $3p_{x,y}$ orbitals. Whereas around $\Lambda$ point VBM consist of Mo $d_{z^2}$ and S $p_z$ and CB of Mo $d_{x^2-y^2, xy}$ and S $p_{x,y}$ and lesser $p_z$ orbital contributions forming hybridized $p$-$d$ bonding and antibonding states respectively.[48,49] On the other hand monolayer ReS$_2$ has a distorted 1T$_d$ structure with VBM and CBM are formed from bonding Re $5d_{x^2-y^2,xy}(t_{2g})$, S $3p$ orbitals primarily $p_z$ and antibonding Re $5d^*_{x^2-y^2,xy}$, S $3p^*$ orbitals[50] with direct band gap of 1.5 eV at Γ point.[18] The Re alloyed MoS$_2$ can exhibit either 1H or 1T$_d$ phase depending on Re concentration as discussed earlier using HRTEM images. Cohesive energy calculation based on fine sampling of composition for two different alloy structures reveal a structural cross over at 37.5% Re concentration which is less than earlier prediction (Fig .6). The stable 1H alloy phases are found to be metallic in nature with Fermi level ($E_f$) shifts into conduction band with no secondary minima at Λ point as observed for pristine MoS$_2$ case. $E_f$ moves further deep into conduction band with increasing Re concentration (Fig.7). The major orbitals contributing near $E_f$ are Re and Mo $d_{z^2}$, $d_{x^2-y^2,xy}$ and S $p$ orbitals (Fig.7(a)-(d)). These orbital contributions are more uniform over $K-\Lambda-\Gamma$ direction compare to MoS$_2$ where weight of each orbital contribution differ along given direction indicate delocalised nature of electron in these alloys.[47]

The alloyed system in 1T$_d$ phase remain metallic till 43.75 Re % (Fig.6(f)-(h)) and a small band gap of 72 meV appears at 50% as reported earlier.[32] These 1T$_d$ phases have same orbital contribution as ReS$_2$ at Fermi level that is Re $d_{x^2-y^2,xy}$ in addition to Mo orbitals. These 1T$_d$ alloys exhibit flat band dispersion. In ReS$_2$, Re atom tetramerization is formed by Re-Re atom hybridisation involving $5d_{x^2-y^2,xy}$ orbitals. The similar bonding nature is observed in Re alloyed 1T$_d$ phases with Re-Re, Mo-Mo and Mo-Re bonds are formed by $d_{x^2-y^2,xy}$ type of orbitals.

In the present investigation a range of structural variation is observed due to the variation in Re concentrations. From discrete dimer to tetramer and with increasing Re concentration super-structure formation with increasing asymmetry of hybridization. In our earlier report we have stated that 1H to 1T structural cross over at $x = 50\%$ from the theoretical calculation. However, fine sampling of composition points reveals that the cross over occurs at ~ 37.5% (Fig. 6). We see diverse interactions between cations, this suggests that the hybridization is not only expected Re-Re but also between Re-Mo and different combinations of these.

## V. Conclusions

In conclusion, we have shown rich diversity of cation hybridization leading to variety of super-structural modulation in $Mo_{1-x}Re_xS$ system for $x < 50\%$. For $x < 23\%$, only short-range interaction is observed. For $x > 23\%$ super-structural modulation is observed, and asymmetry of modulation vector increases with Re concentration. Theoretical calculation suggests that interaction between cations is not only limited between Re atoms but also between Re-Mo atoms.

## ACKNOLEDGMENT

The authors sincerely thank to ICMS, JNCASR for the financial support.

# TABLES

**TABLE I.** Average lattice parameters of Re$_x$Mo$_{1-x}$S$_2$ alloy in Å for different Re concentrations.

| Re % | $a_1$ | $a_1'$ | $a_2$ | $a_2'$ | $a_3$ | $a_3'$ | $\alpha_1$ | $\alpha_2$ |
|---|---|---|---|---|---|---|---|---|
| 100 | 2.7 | 3.3 | 2.5 | 3.5 | 2.5 | 3.3 | 59 | 62 |
| 23 | 2.88 | 2.95 | 3.01 | 3.07 | 3.17 | 3.19 | 62 | 55 |
| 32 | 2.83 | 3.13 | 2.75 | 3.48 | 2.58 | 3.06 | 54 | 63 |
| 38 | 2.57 | 2.66 | 2.68 | 3.35 | 2.73 | 3.64 | 63 | 56 |
| 45 | 2.37 | 2.37 | 2.34 | 3.69 | 2.34 | 3.69 | 59 | 60 |

**TABLE II.** Angles between tree modulation vectors in FFT in degree and its deviation from hexagonal symmetry $\delta\beta_n = |\beta_n - 60|$ for different Re concentration.

| Re % | $\beta_1$ | $\beta_2$ | $\beta_3$ | $\delta\beta_1$ | $\delta\beta_2$ | $\delta\beta_3$ |
|---|---|---|---|---|---|---|
| 0 | 60 | 60 | 60 | | | |
| 23 | 61 | 62 | 57 | 1 | 1 | 3 |
| 32 | 56 | 63 | 61 | 4 | 3 | 1 |
| 38 | 69 | 61 | 50 | 9 | 1 | 10 |
| 45 | 68 | 68 | 44 | 8 | 8 | 16 |
| 100 | 55 | 64 | 61 | 5 | 4 | 1 |

**TABLE III.** Cohesive energies of various structures for Re$_x$Mo$_{1-x}$S$_2$ alloy with $x = 25\%$ considered for theoretical calculations. Relative energy difference between the structures is also given.

| Mo$_{0.75}$Re$_{0.25}$S$_2$ | Cohesive energy (eV) | $\Delta E = E_m - E$ meV | Band Gap (eV) |
|---|---|---|---|
| a | -5.25249 |  | Metallic |
| b | -5.24964 | -2.850 | 0.050 |
| c | -5.23752 | -14.97 | 0.123 |
| d | -5.23691 | -15.58 | Metallic |
| e | -5.06348 | -189.01 | Metallic |

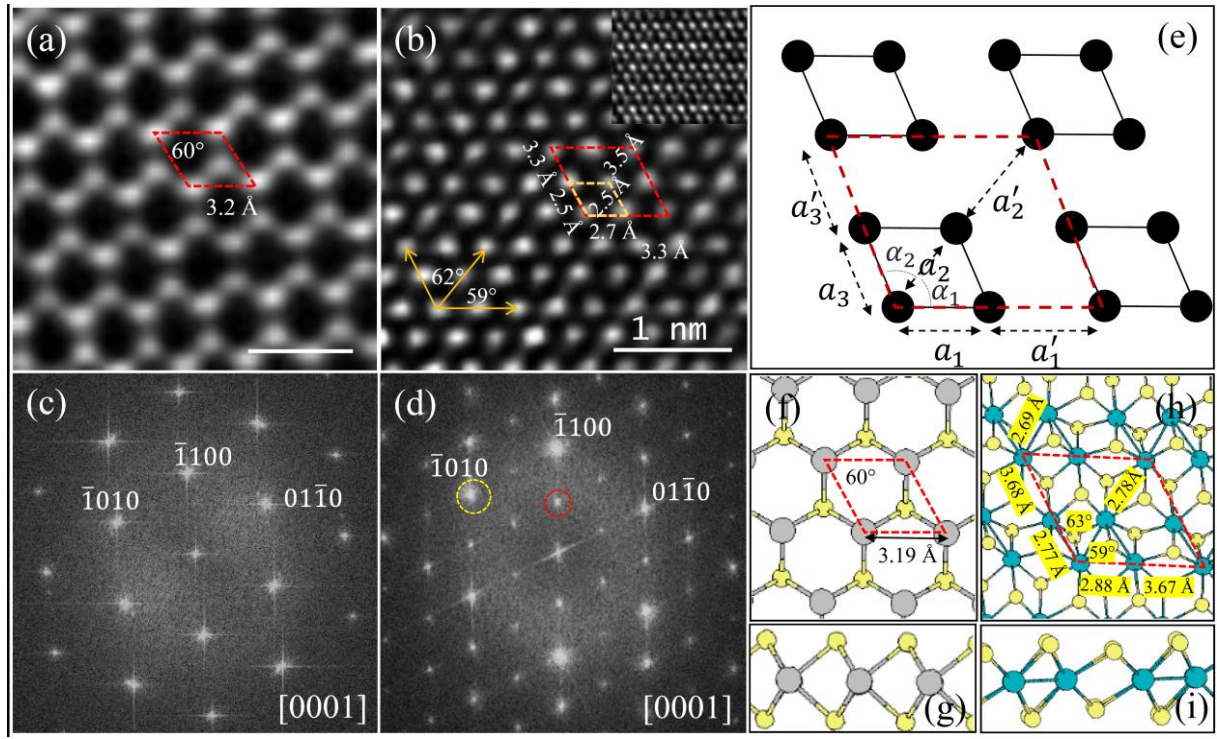

**FIG. 1.** (a) & (b) are the HRTEM images of MoS$_2$ and ReS$_2$ with corresponding FFT is given below it in (c) & (d), respectively. In (b) the inset on right corner shown clear $2a \times 2a$ supercell of ReS$_2$ (d) In ReS$_2$ FFT superlattice and Bragg spot is marked by red and yellow circle. (e) Schematic supercell with lattice parameters indicated which are used to describe the observed super-structure in the text. (f) & (h) are the schematic structure of MoS$_2$ and ReS$_2$ with lattice parameters obtained from the DFT calculation, respectively. (g) Side view of MoS$_2$ depicting trigonal prismatic geometry and (h) side view of ReS$_2$ with showing distorted octahedral geometry.

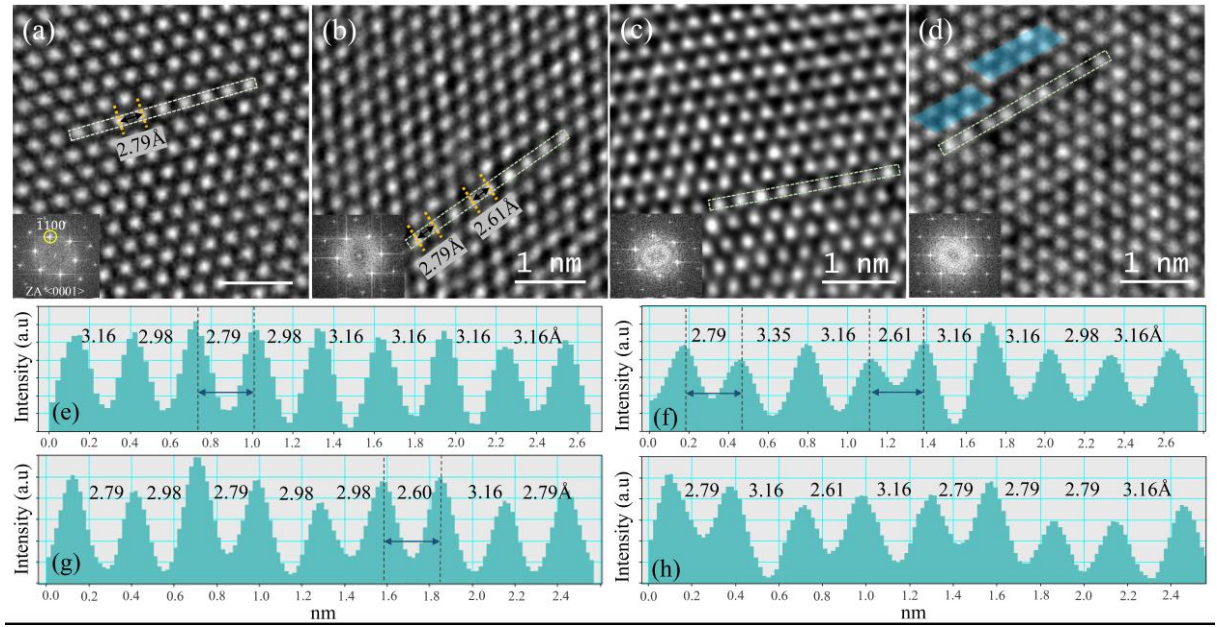

**FIG. 2.** (a)-(d) HRTEM images of $Re_xMo_{1-x}S_2$ alloy for $x = 1, 7, 14, \& 17\%$. The corresponding FFT pattern is given in the inset showing absence of any super-lattice spots. (e)-(h) are the line profiles from each image, respectively from the green dotted box showing the distribution of inter-cation distance. Shorter inter cation bond length is indicated with black arrow. The number of dimers increase in (b) compared to (a). In (d) local tetramer cluster is marked by blue parallelogram.

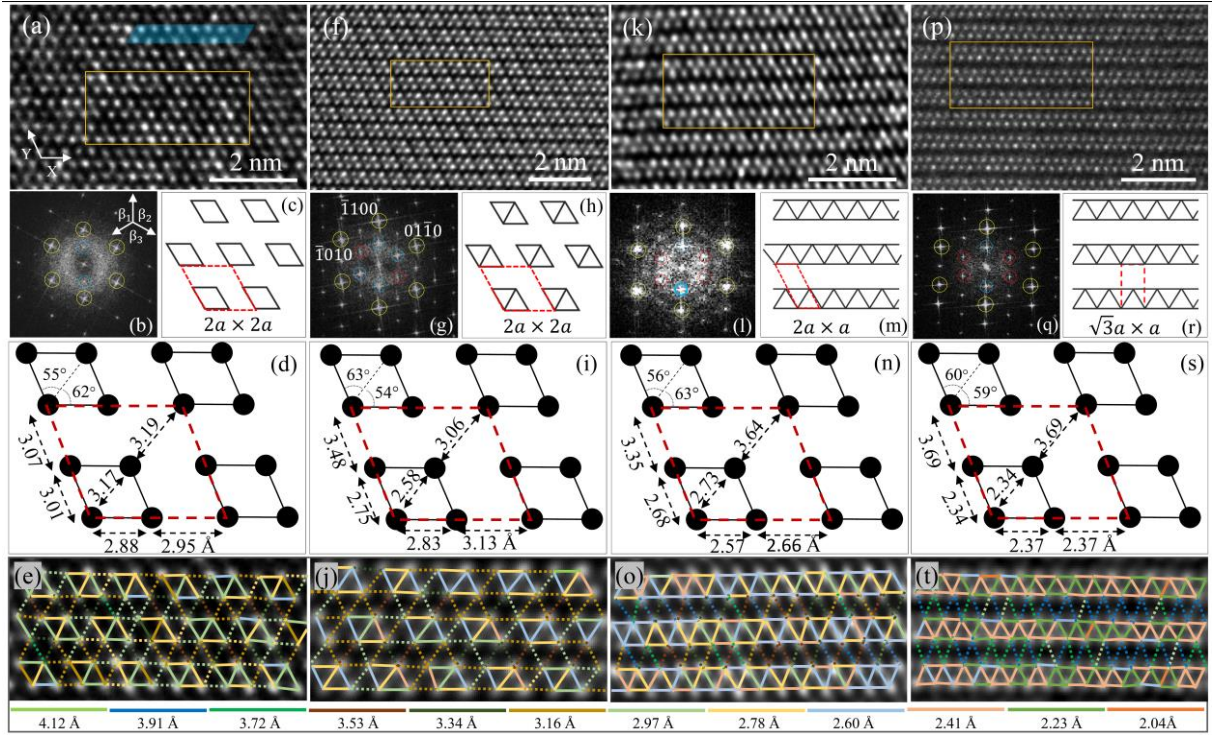

**FIG. 3.** (a)-(d) HRTEM image showing various super-structure modulation $x = 23, 32, 38,$ and $45\%$ with selected area marked by yellow box to show cation-cation distances with different magnitude in (e), (j), (o), and (t), respectively. Below each HRTEM image, FFT pattern showing super-lattice spots, schematic of cation interaction, periodic supercell and average lattice parameters are given. In FFT, usual spots are marked by yellow circles, strong superlattice spots are marked by doted blue circles and weak spots by red circles.

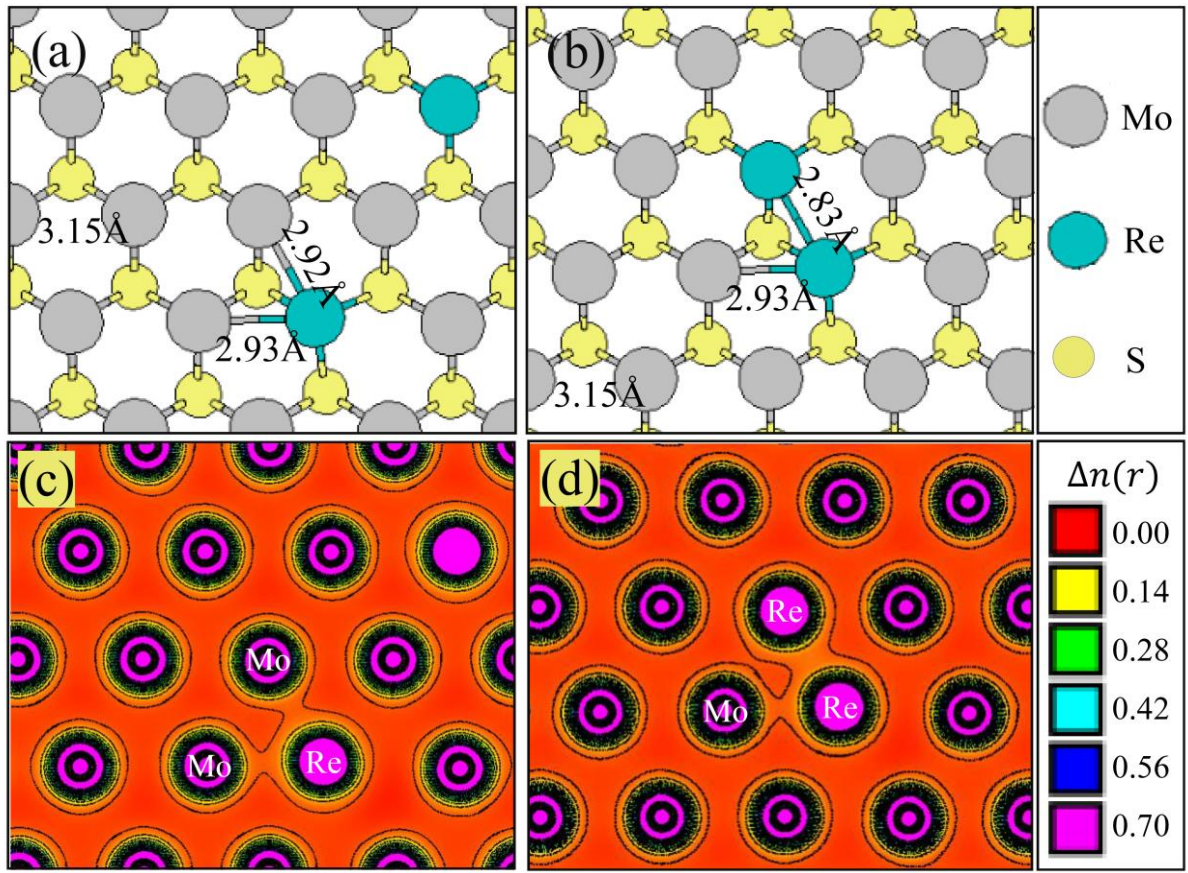

**FIG. 4.** (a) & (b) Schematic structure of Re$_x$Mo$_{1-x}$S$_2$ alloy for $x = 12.5\%$ in 2H configuration for two different Re distributions and cation distances are mentioned in Å. Bottom panel is the charge density plot along <0001> direction, showing the interaction between Re and Mo atoms.

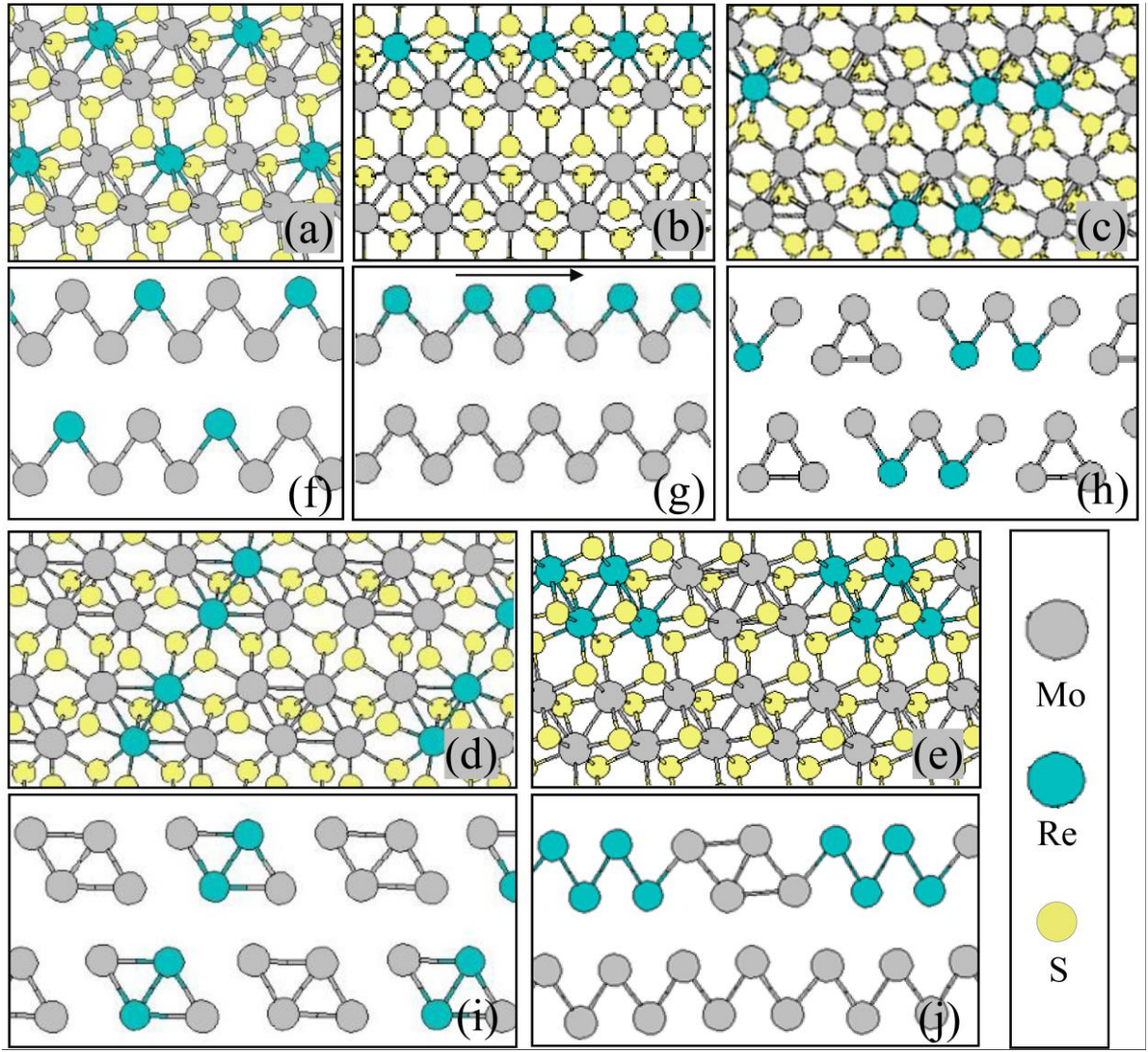

**FIG. 5.** (a)-(e) are schematic structure of $Re_xMo_{1-x}S_2$ alloy for $x = 25\%$ in $1T_d$ configuration considered for DFT calculation with decreasing stability. The bottoms schematics show only cation in the lattice with various inter-cation interactions.

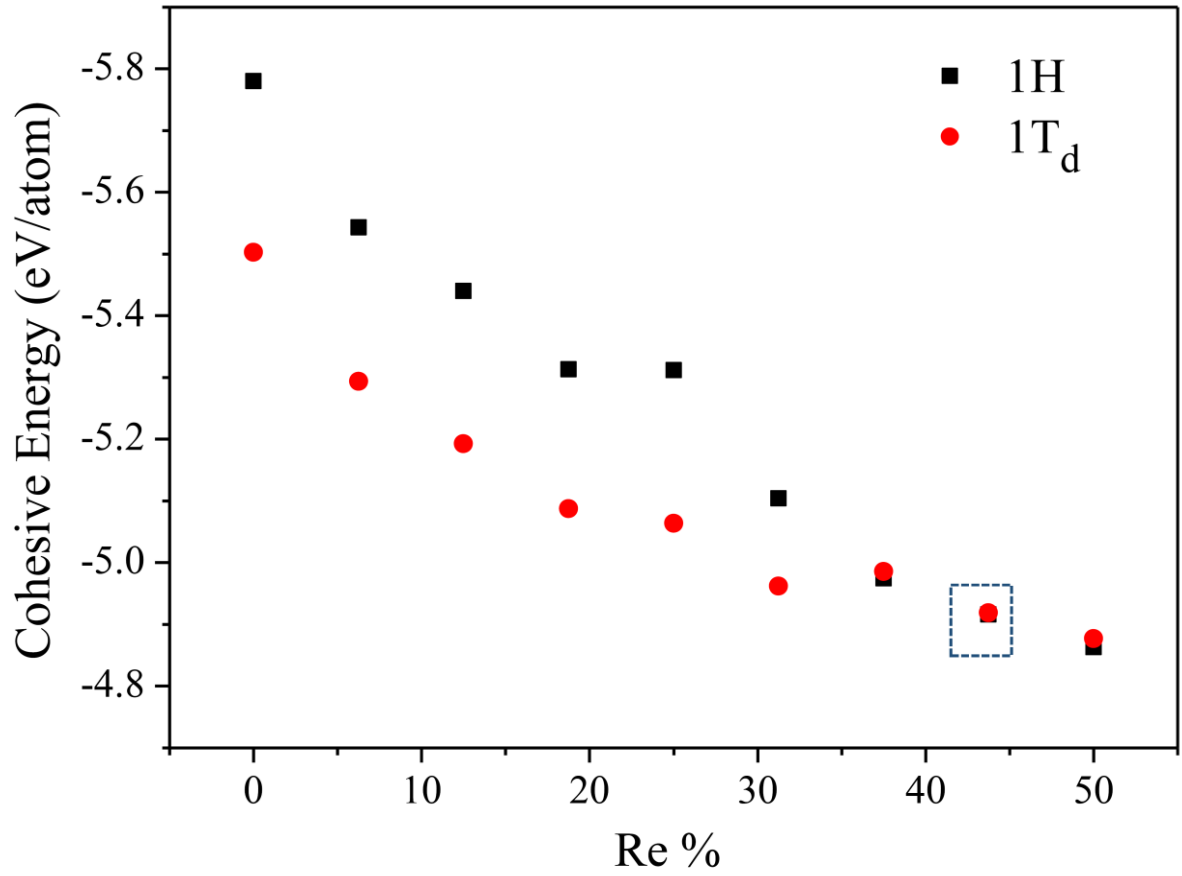

**FIG. 6.** The cohesive energy of 1H and $1T_d$ phase for different Re concentration is shown with cross over is found to be at $x = 37.5\%$ marked by grey box.

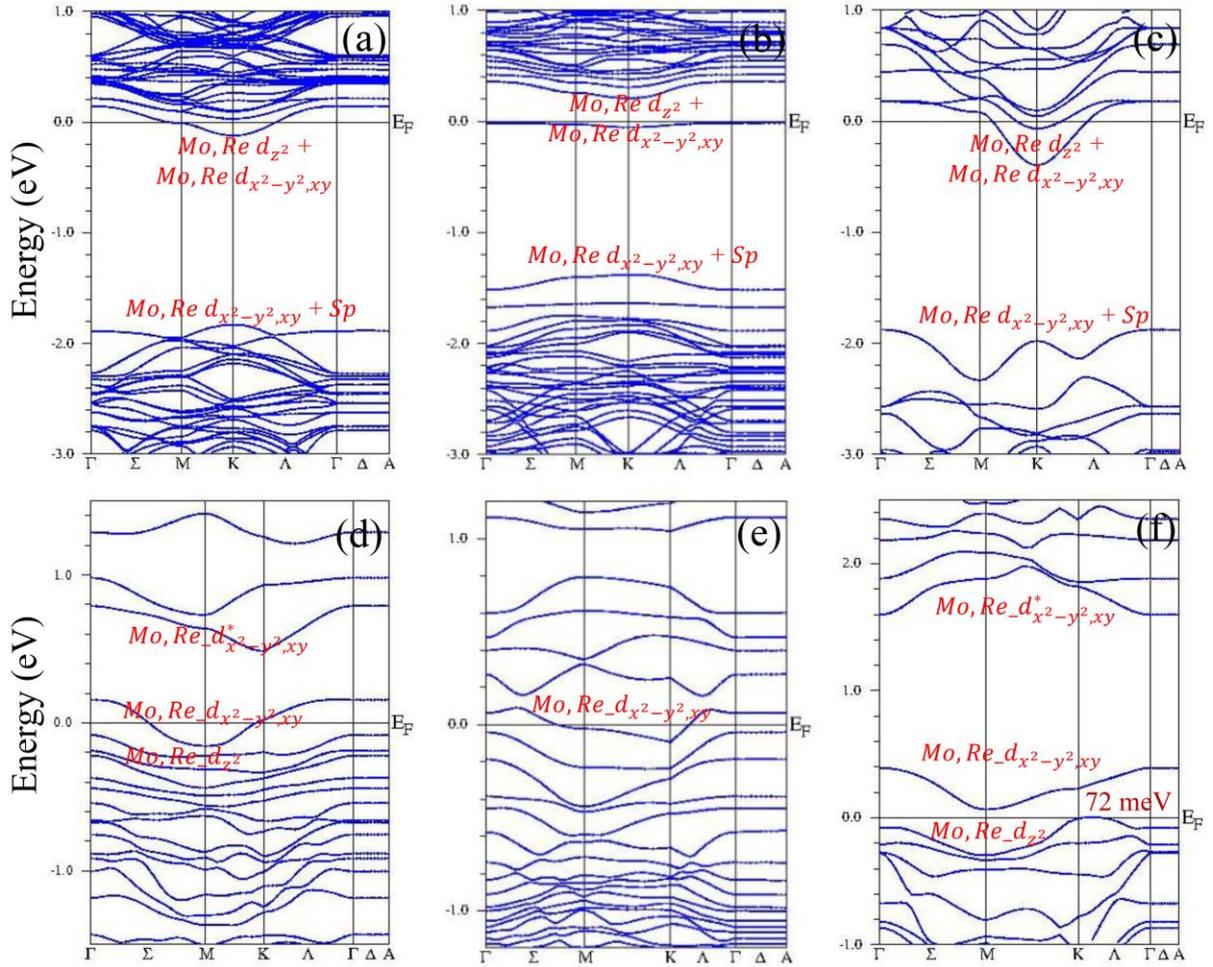

**FIG.7.** Band structure of 1H alloy case for Re concentration (a) 6.25%, (b) 12.5% and (c) 25%. The band structure of $1T_d$ alloyed system for Re concentration (d) 37.50%, (e) 43.50% and (f) 50% with the orbitals contributing near fermi level are mentioned for each case.